\documentclass[aps,prl,reprint,superscriptaddress]{revtex4-1}

\usepackage[hidelinks]{hyperref}
\newcounter{myequation}
\newcounter{myfigure}
\newcounter{mytable}
\makeatletter
\@addtoreset{equation}{myequation}
\@addtoreset{figure}{myfigure}
\@addtoreset{table}{mytable}
\makeatother
\usepackage{graphicx}
\usepackage{bm}
\usepackage{amsmath}
\usepackage{amssymb}
\usepackage{todonotes}
\usepackage{physics}
\usepackage{siunitx}

\newcommand{\kv}{\bm{k}}
\newcommand{\rv}{\bm{r}}
\newcommand{\Rv}{\bm{R}}
\newcommand{\phiM}{\phi_M}
\newcommand{\phiH}{\phi_H}
\newcommand{\psik}{\bm{\psi}_{\kv}}
\newcommand{\Ilead}{I^{\alpha}_L}

\DeclareMathOperator{\diag}{diag}

\newcommand{\gv}{\bm{g}}

\newcommand{\nv}{\bm{S}}
\newcommand{\av}{\hat{\bm{a}}}
\newcommand{\bv}{\hat{\bm{b}}}
\newcommand{\xv}{\hat{\bm{x}}}
\newcommand{\expn}{\hat{\bm{n}}}

\DeclareMathOperator{\sig}{sig}
\DeclareMathOperator{\erfc}{Erfc}

\begin{document}
	
	
	\title{Electrical detection of unconventional transverse spin-currents in obliquely magnetized thin films}
	
	
	\author{Pieter M. Gunnink}
	\email{p.m.gunnink@uu.nl}
	\affiliation{Institute for Theoretical Physics and Center for Extreme Matter and Emergent Phenomena, Utrecht University, Leuvenlaan 4, 3584 CE Utrecht, The Netherlands}
	
	
	\author{Rembert A. Duine}
	\affiliation{Institute for Theoretical Physics and Center for Extreme Matter and Emergent Phenomena, Utrecht University, Leuvenlaan 4, 3584 CE Utrecht, The Netherlands}
	\affiliation{Department of Applied Physics, Eindhoven University of Technology, P.O. Box 513, 5600 MB Eindhoven, The Netherlands}
	\author{Andreas R\"uckriegel}
	\affiliation{Institute for Theoretical Physics and Center for Extreme Matter and Emergent Phenomena, Utrecht University, Leuvenlaan 4, 3584 CE Utrecht, The Netherlands}
	
	\date{\today}
	
	\begin{abstract}
		In a typical experiment in magnonics, thin films are magnetized in-plane and spin waves only carry angular momentum along their spatial propagation direction. Motivated by the experiments of Bozhko \textit{et al.} [Phys. Rev. Research 2, 023324 (2020)], we show theoretically that for obliquely magnetized thin films, exchange-dipolar spin waves are accompanied by a transverse spin-current. We propose an experiment to electrically detect this transverse spin-current with Pt strips on top of a YIG film, by comparing the induced spin-current for spin waves with opposite momenta. We predict the relative difference to be of the order $10^{-4}$, for magnetic fields tilted at least $30^{\circ}$ out of plane. This transverse spin-current is the result of the long range dipole-dipole interaction and the inversion symmetry breaking of the interface.
	\end{abstract}

	\maketitle

	\paragraph{Introduction.}
	Magnons, or spin waves, are able to transport angular momentum over long distances along their propagation direction \cite{cornelissenLongdistanceTransportMagnon2015,gilesLongrangePureMagnon2015}. This has opened the way to novel signal processing devices which could replace conventional electronic devices \cite{stamps2014MagnetismRoadmap2014,csabaPerspectivesUsingSpin2017,klinglerSpinwaveLogicDevices2015}. In recent years, multiple applications have been explored, such as wave-based computing \cite{khitunMagnonicLogicCircuits2010, khitunNonvolatileMagnonicLogic2011}, three-terminal transistors \cite{chumakMagnonTransistorAllmagnon2014}, logic gates \cite{schneiderRealizationSpinwaveLogic2008, fischerExperimentalPrototypeSpinwave2017} and novel non-linear effects \cite{melkovWaveFrontReversal2000, sadovnikovNonlinearMagnonicsIntensitydependent2017}.

	
	The manipulation of spin waves is still an ongoing area of research and a full toolbox for controlling spin waves is yet to be developed \cite{chumakMagnonSpintronics2015}. In this work we consider an alternative approach to control the spin current in a magnetized thin film: by tilting the magnetic field out of plane. This breaks the inversion symmetry and allows a spin current to flow transverse to the propagation direction of the spin waves, transporting angular momentum along the film normal.

	This mechanism for generating a transverse spin-current was first proposed by \textcite{bozhkoUnconventionalSpinCurrents2020}, who used a micromagnetic approach to calculate the exchange spin-current in a thin film of Y\textsubscript{3}Fe\textsubscript{2}(FeO\textsubscript{4})\textsubscript{3} (YIG), without considering spin absorption at the boundaries. They argued that this spin current is non-zero if the magnetic field is tilted out of plane. However, this transverse spin-current can only be detected with an attached spin sink, such as a heavy metal strip. The interaction with the spin sink influences the physics of the problem significantly. Moreover, only the transfer of angular momentum by the exchange interaction was considered. The dipole-dipole interaction is also capable of transporting angular momentum and therefore needs to be taken into account for a complete description of this system.
	
	
	
	In this work we propose an experiment where the transverse spin-current in an obliquely magnetized thin film is detected electrically. We consider, within linear spin-wave theory, a thin ferromagnetic film with two leads attached, which pick up the transverse spin-current induced by left- and right-moving spin waves via the inverse spin-Hall effect (ISHE) \cite{saitohConversionSpinCurrent2006}. A transverse spin-current would transport more angular momentum into the right spin sink than into the left spin sink, or vice versa. This is equivalent to the experimentally harder to realize system with leads attached to the top and bottom. We propose to compare the spin current picked up by the left and right lead, in order to exclude any usual spin pumping effects, which are also present for an in-plane magnetic field \cite{tserkovnyakEnhancedGilbertDamping2002}. In order to further understand the origin of the transverse spin-current we show in the supplemental material \footnote{See Supplemental Material for the full form of the amplitude factors in Eq.~(\ref{eq:d-matrix}), the dispersions, the details of the magnetostatic calculation and the explicit form of the terms in the continuity equation for the spin in Eq.~(\ref{eq:continuity}).} with a magnetostatic calculation, that the symmetry breaking at the interface is carried by the dipole-dipole interaction.

\begin{figure}
		\centering
		\includegraphics[width=\columnwidth]{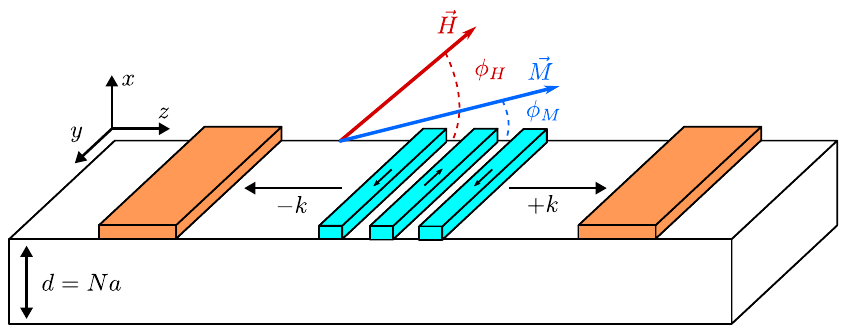}
		\caption{The setup considered, with a coplanar waveguide in the middle, exciting spin waves in two opposite directions in a thin ferromagnetic film with thickness $d$. Two heavy-metal leads pick up the spin current induced by these left- and right-moving spin waves. The magnetic field is tilted out of plane at an angle $\phiH$ with the plane and the magnetization has angle $\phiM$ with the plane.
			\label{fig:setup}}
	\end{figure}

	\paragraph{Method.}
	The setup we consider is a thin film of ferromagnetic YIG, where coherent spin-waves are excited using a coplanar waveguide \cite{fallarinoPropagationSpinWaves2013}, as depicted in Fig.~\ref{fig:setup}. The wavevector ($k$) of the excited magnons is controlled by the grating of the antenna and the frequency ($\omega$) of the excited magnons by the frequency of the driving field. To the right and left of this antenna two platinum (Pt) leads are placed which function as spin sinks via the inverse spin-Hall effect and pick up the transverse spin-current induced by the spin waves with opposite momenta. The distance between the Pt leads and the coplanar waveguide is assumed to be such that the signal is strong enough to measure small variations. Structures with a separation distance of \SI{3}{mm} are possible \cite{chumakDirectDetectionMagnon2012}, but the magnon diffusion length of $\lambda=\SI{9.4}{\mu m}$ in YIG \cite{cornelissenLongdistanceTransportMagnon2015} indicates that shorter distances would be preferable.
	
	The spin dynamics are governed by the semi-classical Landau-Lifshitz-Gilbert (LLG) equation:
	\begin{equation}
	\partial_{t}\bm{S}_{i}
	= \bm{S}_{i}\times\left(-\frac{\partial \mathcal{H}}{\partial\bm{S}_{i}}+\bm{h}_{i}(t)
	-\frac{\alpha_i}{S}  \partial_{t}\bm{S}_{i}\right),
	\label{eq:LLG}
	\end{equation}
	where we describe YIG as a Heisenberg ferromagnet with effective spin $S$, on a cubic lattice. Including both the exchange and dipole-dipole interactions our effective Hamiltonian \cite{cherepanovSagaYIGSpectra1993} is 
	\begin{multline}
	\mathcal{H}=
	-\frac{1}{2}\sum_{ij}J_{ij}\bm{S}_{i} \cdot \bm{S}_{j}-\mu\bm{H}_{e}\cdot\sum_{i}\bm{S}_{i}
	\\ -\frac{1}{2}\sum_{ij,i\neq j}\frac{\mu^2}{|\bm{R}_{ij}|^{3}}\left[3\left(\bm{S}_{i}\cdot\hat{\bm{R}}_{ij}\right)\left(\bm{S}_{j}\cdot\hat{\bm{R}}_{ij}\right)-\bm{S}_{i}\cdot\bm{S}_{j}\right],
	\label{eq:ham-short}
	\end{multline}
	where the sums are over the lattice sites $\bm{R}_{i}$, with $\bm{R}_{ij}=\bm{R}_i-\bm{R}_j$ and $\hat{\bm{R}}_{ij} = \bm{R}_{ij} / |\bm{R}_{ij}|$. We only consider nearest neighbour exchange interactions, so $J_{ij}=J$ for nearest neighbours and 0 otherwise. Here $\mu = 2\mu_B$ is the magnetic moment of the spins, with $\mu_B=e\hbar/(2m_ec)$ the Bohr magneton. $\bm{H}_{e}$ is the external magnetic field, which we take strong enough to fully saturate the ferromagnet.
	
	To the top of the thin film we attach a spin sink to detect the spin waves, which introduces an interfacial Gilbert damping $\alpha^L_i$, which is only non-zero for sites at the top interface of the ferromagnet \cite{tserkovnyakEnhancedGilbertDamping2002}. The total Gilbert damping is then $\alpha_i=\alpha^B+\alpha^L_i$, where $\alpha^B$ is the bulk Gilbert damping. Furthermore, $\bm{h}_i(t)$ is the circularly polarized driving field, which we take to be uniform throughout the film. Within linear spin-wave theory, the LLG has been shown to be fully equivalent to the non-equilibrium Green’s function formalism \cite{zhengGreenFunctionFormalism2017}.
	
	We consider a thin film, infinitely long in the $y,z$ directions and with a thickness $d=Na$ in the $x$ direction, where $a$ is the lattice constant and $N$ is the number of layers. The magnetic field is tilted at an angle $\phi_H$ with respect to the film, as shown in Fig.~\ref{fig:setup}. The magnetization is tilted by an angle $\phi_M$, as determined by minimizing the energy given by Eq.~(\ref{eq:ham-short}) for a classical, uniform spin configuration:
	\begin{equation}
	\frac{\partial}{\partial\phi_{M}}\left[-M_sH_e\cos\left(\phi_M-\phi_H\right)-2\pi M_s^{2}\cos^{2}\phi_H\right]=0,
	\label{eq:angle-magnetic-field}
	\end{equation}
	where $M_s=\mu S/a^3$ is the saturation magnetization and $H_e=|\bm{H}_e|$.
	
	We have two reference frames, one aligned with the thin film as described above and one where the $z$ axis is aligned with the magnetization $\bm{M}$. We work in the reference frame of the lattice and rotate the spin operators, such that $\bm{S}_i \rightarrow {\cal R}_y^{-1}\left(\phiM\right) \bar{\bm{S}}_i$, 
	where ${\cal R}_y\left(\phiM\right)$ is a rotation around the $y$-axis by angle $\phiM$ and $\bar{\bm{S}}_i$ are the rotated spin operators, with the $\bar{S}^z_i$ component pointing along the magnetization $\bm{M}$. 
	
	We linearize in the deviations from the ground state, $b_i=\frac{1}{2}\sqrt{2S} \left(\bar{S}^x_i + i \bar{S}_i^y\right)$ and assume translational invariance in the $y$$z$-plane. The equation of motion for $b_i$ becomes  in frequency space:
	\begin{equation}
	\mathbb{G}^{-1}_{\kv}(\omega)\psik(\omega) = -\bm{h}_{\kv}(\omega),	\label{eq:greens}
	\end{equation}
	where $\kv=(k_y, k_z)$ and we have introduced the driving field
	\begin{equation}
	\bm{h}_{\kv}(\omega) = (
	\underbrace{h_{\kv}(\omega),...,h_{\kv}(\omega)}_{N\text{ elements}}, \underbrace{h^*_{-\kv}(\omega),...,h^*_{-\kv}(\omega)}_{N\text{ elements}}
	)^T,
	\end{equation} 
	where $h_{\kv}(\omega)=\bar{h}_x+i\bar{h}_y$ is the Fourier transform of the rotated driving field. Furthermore, the magnon state vector is
	\begin{multline}
	\qquad \psik(\omega)=\big(
	b_{\kv}(\omega, x_1),...,b_{\kv}(\omega, x_N), \\
	b_{-\kv}^*(\omega, x_1),...,b^*_{-\kv}(\omega, x_N)
	\big)^T
	\end{multline}
	and the inverse Green's function is
	\begin{equation}
	\mathbb{G}^{-1}_{\kv}(\omega) = \sigma_{3}  \left(1+i\sigma_{3}{\alpha}\right)\omega - \sigma_3 \mathcal{H}_{\kv},
	\end{equation}
	where we have introduced $\sigma_3=\diag\left(1,...,1,-1,...,-1 \right)$, ${\alpha} = \diag\left(\alpha_1,...,\alpha_N,\alpha_1,...,\alpha_N\right)$ and
	\begin{equation}
	\mathcal{H}_{\kv} = \begin{pmatrix}
	\bm{A}_{\kv} & \bm{B}_{\kv}\\
	\bm{B}_{\kv}^{\dagger} & \bm{A}_{\kv}
	\end{pmatrix},
	\label{eq:d-matrix}
	\end{equation}
	which is the Hamiltonian matrix within linear spin-wave theory, with the amplitude factors $\left[\bm{A}_{\kv}\right]_{ij} = {A}_{\kv}\left( x_i-x_j \right)$ and $\left[\bm{B}_{\kv}\right]_{ij} = {B}_{\kv}\left(x_i-x_j\right)$. The dispersion is obtained by diagonalizing the inverse Green's function~(\ref{eq:greens}) in the absence of damping and spin pumping. The full expressions for the amplitude factors $A_{\kv}, B_{\kv}$ and the dispersions for different tilting angles of the magnetic field are given in the supplemental material \cite{Note1}.

	From the equation of motion, Eq.~(\ref{eq:greens}), the total spin-current injected into the lead is obtained from the continuity equation for the spin:
	\begin{equation}
	\partial_t \bar{S}_{i}^{z}+\sum_{j}I^{ex}_{i\rightarrow j}+\sum_{j}I^{dip-dip}_{i\rightarrow j}=I_{i}^{\alpha}+I_{i}^{h}.
	\label{eq:continuity}
	\end{equation}
	The explicit form of the terms is given in the supplemental material \cite{Note1}. We find a source and sink term, providing angular momentum via the driving field ($I_{i}^{h}$) and dissipating angular momentum to the lattice and the lead via the Gilbert damping ($I_{i}^{\alpha}$). There are two ways angular momentum can be transferred through the film. Firstly, there is a spin current transferring angular momentum between adjacent sites ($I^{ex}_{i\rightarrow j}$), which is driven by the exchange interaction. 
	The dipole-dipole interaction also transports angular momentum ($I^{dip-dip}_{i\rightarrow j}$), but because the dipole-dipole interaction is non-local, angular momentum is transferred from and to all other sites. It is therefore not possible to write this as a local divergence and thus as a current. Also note that the dipole-dipole interaction couples the magnons to the lattice, which means that a non-zero dipole-dipole contribution is accompanied by a transfer of angular momentum from and to the lattice. 
	
	The measurable quantity is the angular momentum absorbed by the spin sink in the attached lead, which is proportional to the voltage generated by the ISHE, and is given by
	\begin{equation}
	\Ilead(\kv, \omega) = 2\alpha^L\Im\left[b_{\kv}^* (x_1) \partial_t b_{\kv} (x_1)\right].
	\label{eq:spincurrentlead}
	\end{equation}
	We are interested in the relative difference between the spin currents induced by the left- and right-moving spin waves in order to show a transverse spin transport, which we define as
	\begin{equation}
	\Delta(|\kv|, \omega) = \frac{\Ilead(\kv, \omega)-\Ilead(-\kv, \omega)}{\max\left[|\Ilead(\kv, \omega)|,|\Ilead(-\kv, \omega)|\right]}.
	\label{eq:delta}
	\end{equation}
	In the next section we consider this quantity in detail.
	
	\begin{table}[]
		\centering
		\caption{Parameters for YIG used in the numerical calculations in this work. Note that $S$ follows from $S=M_sa^3/\mu$.\label{tab:values}}
		\begin{ruledtabular}
			\begin{tabular}{@{\hspace{6em}} lc @{\hspace{6em}}}
				Quantity & Value \rule{0pt}{2.6ex}\rule[-1.2ex]{0pt}{0pt}\\
				\hline
				$N$ & $400$ \rule{0pt}{2.6ex}\\
				$a$ &  $\SI{12.376}{\angstrom}$  \cite{gellerCrystalStructureFerrimagnetism1957}  \\
				$S$ &  14.2 \\
				$4\pi M_s$ & $\SI{1750}{G}$  \cite{tittmannPossibleIdentificationMagnetostatic1973}  \\
				$J$ & $\SI{1.60}{K}$ \cite{kreiselMicroscopicSpinwaveTheory2009} \\
				$\alpha^B$ & $\num{7e-4}$ \cite{haertingerSpinPumpingYIG2015} \\
				$\alpha^L$ & $\num{7e-3}$ \cite{haertingerSpinPumpingYIG2015}  \\
				$H_e$ & $\SI{2500}{Oe}$ \\
				$h_x, h_y$ & $0.01H_e$ 		
			\end{tabular}
		\end{ruledtabular}
	\end{table}
	
\begin{figure*}
		\centering
		\includegraphics[width=\textwidth]{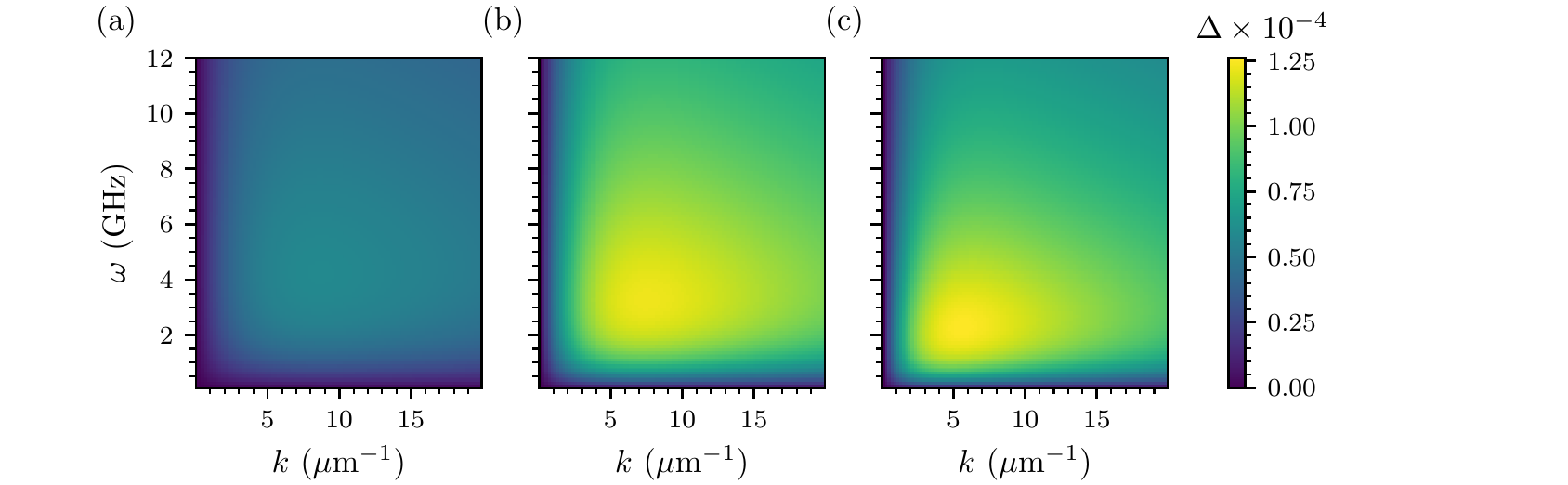}
		\caption{The relative difference $\Delta$ between the spin current induced by left- and right-moving spin waves, as defined in Eq.~(\ref{eq:delta}), as a function of $k$ and $\omega$, for three different tilting angles of the magnetic field. The spin waves travel parallel to the in-plane projection of the magnetic field, such that $\kv=k \hat{\bm{z}}$. (a) $\phiH=\ang{30}, \phiM=\ang{18}$, (b) $\phiH=\ang{60}, \phiM=\ang{40}$ and (c) $\phiH=\ang{80}, \phiM=\ang{64}$. The peak difference is $\Delta(k=\SI{7.5}{\micro \meter^{-1}}, \omega=\SI{4}{GHz})=\num{1.25e-4}$, when the field is tilted at an angle $\phiH=\ang{60}$. For a magnetic field completely in- or out of plane (not shown) there is no discernible difference.
			\label{fig:heatmap}}
	\end{figure*}

	\paragraph{Results.}
	The parameters used throughout this work are summarized in Table.~\ref{tab:values}. In Fig.~\ref{fig:heatmap} we show the difference between the spin current induced by left- and right-moving spin waves for different tilting angles of the magnetic field. For a magnetic field either completely in- or out of plane there is no difference between the left and right lead (not shown). As we tilt the magnetic field out of plane a small difference becomes visible, which peaks at $\Delta=1.25\times10^{-4}$ for $\phiH=\ang{60}$ and $2.5<k<\SI{12.5}{\mu m^{-1}}$. As the tilting angle is further increased the distribution of $\Delta$ shifts slightly, with the most notable change the movement of the maximum, which moves towards smaller wavevectors. We found that the relative difference $\Delta$ increases linearly with the bulk Gilbert damping constant. In order to measure this effect it might therefore be beneficial to use a YIG thin film with deliberately introduced impurities such as rare-earth ions, to increase the damping \cite{sharmaMagneticCrystallographicProperties2018}, or even use a different ferromagnetic material with a higher Gilbert damping.
	
	Numerically, we found that the relative difference $\Delta$ is non-zero even when the exchange coupling is artificially turned off, which indicates that only the dipole-dipole interaction is responsible for this effect. In the supplemental material \cite{Note1} we show a full magnetostatic derivation of the eigenmodes for an obliquely magnetized thin film with only dipole-dipole interactions. Even though the energies are inversion-symmetric, we find that the eigenmodes explicitly depend on
	\begin{equation}
	k_z\sin\left(2\phiM\right),
	\end{equation}
	which introduces an asymmetry between left- and right-moving spin waves if the magnetic field is tilted out of plane. 
	A complete description of this problem also requires the inclusion of the exchange coupling, as was done in our numerical calculations. However, ignoring the exchange coupling allows us to demonstrate that the origin of the asymmetry between left- and right-moving spin waves lies in the the long range dipole-dipole interaction carrying the inversion symmetry breaking of the interface.

	\textcite{bozhkoUnconventionalSpinCurrents2020} suggested a partial-wave picture to explain the transverse spin-current. They reason that the profile along the film normal is made up by two partial waves, which have opposite momenta $\pm k_x$ and equal frequency $\omega$ if the film is magnetized in-plane, thus cancelling any transfer of angular momentum or energy. As the magnetic field is tilted out of plane the two partial waves would, in this picture, no longer have opposite momenta, but still have the same frequencies. This would then allow for angular momentum transfer, but not energy transfer. With the magnetostatic calculation we are able to show that this picture is incomplete: the amplitudes of the two partial waves are asymmetric, not their momenta. This therefore allows both energy and angular momentum transfer, which we have confirmed numerically by evaluating $\expval{\partial_t E}$.
	
	We found numerically that the region in $k$-space where the relative difference $\Delta$ is significant has a lower bound related to the thickness of the thin film. Decreasing the thickness shifts the distribution as seen in Fig.~\ref{fig:heatmap} towards larger wavevectors. 
	This can be traced to the fact that the long-wavelength magnetostatic magnon modes are standing waves \cite{Note1}, with wavevectors $\pm k_x$, where $k_x$ is proportional to $k_z$. The standing waves need to have a wavevector big enough to fit at least one wavelength into the system, thus requiring that $k_z\gtrsim k_{L}$, where $k_L=2\pi/d$. The reason for this coupling of the in-plane and out of plane directions is the long-range nature of the dipole-dipole interaction, ensuring that within our system the divergence of the magnetic field is zero, i.e., $\nabla\cdot\bm{B}=0$. The maximum value of $\Delta$ does not change depending on the thickness of the film, only the location of the maximum.  We have confirmed this numerically for the range $60\leq d\leq \SI{480}{nm}$. For even thinner films the maximum value of $\Delta$ becomes lower.
	
\begin{figure}
		\centering
		\includegraphics{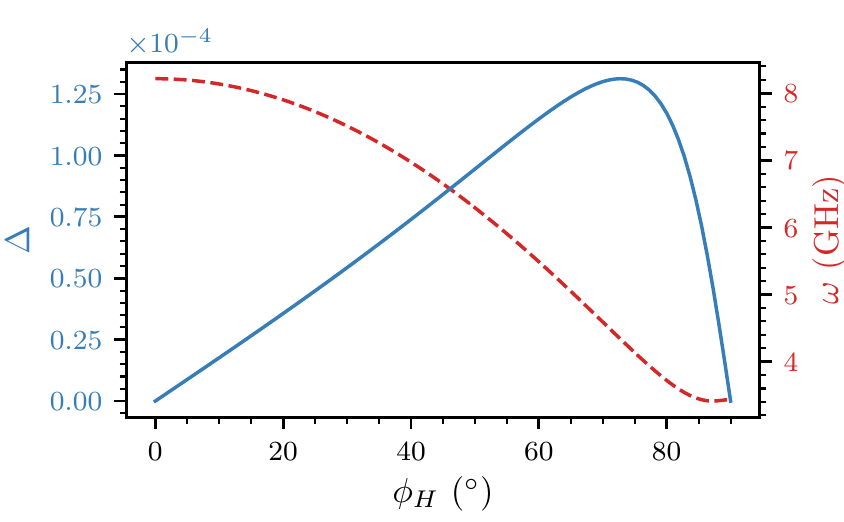}
		\caption{Relative difference between the spin current induced by left- and right-moving spin waves, $\Delta$, as defined in Eq.~(\ref{eq:delta}), as a function of magnetic field tilt angle $\phiH$, for $\omega$ corresponding to the lowest mode in the spin-wave dispersion and fixed $k=\SI{7.5}{\micro \meter^{-1}}$ (solid line). Also shown is the frequency of the lowest mode as a function of the tilt angle (dashed line).
			\label{fig:alpha-scan}}
	\end{figure}

	Excitation of magnons is only possible for values of $\omega$ determined by the spin-wave dispersion, with a minimum given by the lowest mode. We therefore show in Fig.~\ref{fig:alpha-scan} for fixed $k=\SI{7.5}{\micro \meter^{-1}}$ the evolution of the relative spin-current difference $\Delta$ as the magnetic field is tilted out of plane, with a driving at frequency $\omega$ corresponding to the lowest mode in the spin-wave dispersion. Also shown is the frequency of the lowest mode as a function of magnetic field tilt angle. It is clear that up to some critical value of the magnetic field angle $\Delta$ increases linearly, after which it falls off rapidly. It is also clear that the lowest mode is capable of transferring angular momentum along the film normal. This is contrary to the statements made by \textcite{bozhkoUnconventionalSpinCurrents2020}, who predicted that the lowest mode, which has an uniform profile, would not induce a transverse spin-current. This is most likely due to the fact that in their work only the exchange current is considered, whereas we have taken all current contributions into account. Another possible explanation is their expansion in eigenfunctions of the second-order exchange operator, which might have failed to properly take the dipole-dipole interaction into account. 
	
\begin{figure}
		\centering
		\includegraphics{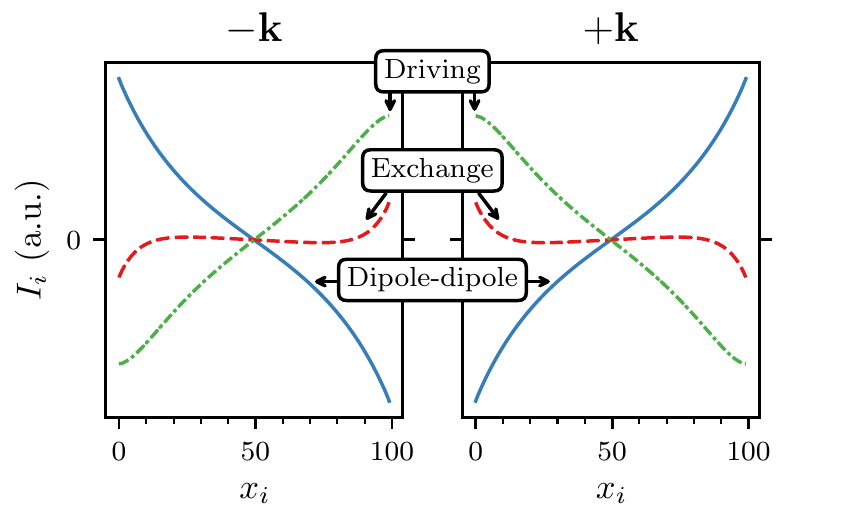}
		\caption{The different contributions to the transfer of angular momentum along the film normal, where $I_i=\sum_j I_{i\to j}$ for the exchange and dipole-dipole interaction. The damping plays a negligible role in the transport of angular momentum, so it is turned off to illustrate the effects of the other contributions. The thickness of the thin film is reduced to $N=100$ in order to better illustrate the variation through the film. The magnetic field is tilted out of plane with angle $\phiM=\ang{60}$ and the wavevector and driving frequency are fixed at $k=\SI{30}{\micro \meter^{-1}}$, $\omega=\SI{4}{GHz}$.  \label{fig:contributions}}
	\end{figure}

	The different contributions to the transverse angular momentum transport, as defined in Eq.~(\ref{eq:continuity}), are shown in Fig.~\ref{fig:contributions} for left- and right-moving spin waves. We have set the bulk and interface damping to zero in order to clearly show the exchange, dipole-dipole and driving contributions to the transfer of spins along the film normal. 
	Firstly, we can see that there is a transport of angular momentum, even in the case of no spin absorption at the boundary, which agrees with the results by \textcite{bozhkoUnconventionalSpinCurrents2020}.
	All contributions are zero in the case of an in-plane magnetic field (not shown)---if no spin sinks are attached. 
	We can see that every contribution switches sign between left- and right-moving spin waves, as would be expected from symmetry.
	From this figure it is clear that the exchange spin current is not the only way the system transfers angular momentum. In fact, the contributions from the dipole-dipole interaction are larger than those of the exchange current. This shows that it is necessary to consider both interactions in order to gain a full understanding of the transport of angular momentum in the transverse direction. Also note that since the dipole-dipole contribution is non-zero there is a finite torque on the system, which could be measured in a cantilever experiment \cite{hariiSpinSeebeckMechanical2019}.

	\paragraph{Conclusion and Discussion.}
	In this work we have shown, using microscopic linear spin-wave theory, that there is a flow of angular momentum, or spin current, along the film normal in obliquely magnetized thin films. This can be measured using an antenna-detector setup, where the spin current induced by the left- and right-moving spin waves will be different, proving the existence of a transverse spin-current. This effect can be used as a way to manipulate the spin current flowing along the film normal, for example by controlling the magnetic field angle. We have also demonstrated that this spin current is the result of the dipole-dipole interactions in the film, which carry the inversion breaking at the interface. 
	
	We have not considered explicitly the interactions of the spin waves with the lattice. The dipole-dipole interactions couple the magnons to the lattice and therefore angular momentum can be transferred from and to the phonons, which can also transport angular momentum \cite{vonsovskiiPhononSpin1962,levineNoteConcerningSpin1962,zhangAngularMomentumPhonons2014}. A more complete description of the system should therefore include these phonon-magnon interactions, but this is beyond the scope of this article.

	\begin{acknowledgments}
		R.D. is member of the D-ITP consortium, a program of the Dutch Organization for Scientific Research (NWO) that is funded by the Dutch Ministry of Education, Culture and Science (OCW). This project has received funding from the European Research Council (ERC) under the European Union’s Horizon 2020 research and innovation programme (grant agreement No. 725509). This work is part of the research programme of the Foundation for Fundamental Research on Matter (FOM), which is part of the Netherlands Organization for Scientific Research (NWO). It is a pleasure to thank Alexander Serga and Huaiyang Yang for discussions.
	\end{acknowledgments}
	
	\bibliography{Unconventional_Spin_Currents}
\onecolumngrid
\pagebreak

\begin{center}
	\textbf{\large Supplemental Material: Electrical detection of unconventional transverse spin currents in obliquely magnetized thin films}
\end{center}

\setcounter{page}{1}
\makeatletter

\stepcounter{myequation}
\stepcounter{myfigure}
\stepcounter{mytable}
\renewcommand{\theequation}{S\arabic{equation}}
\renewcommand{\thefigure}{S\arabic{figure}}
\renewcommand{\bibnumfmt}[1]{[S#1]}

\section{Magnetostatic calculcations}
Our goal is to derive the eigenfunctions for the thin film geometry as depicted in Fig.~\ref{fig:setup} in the main text. We know from the numerics that the dipole-dipole interaction alone is sufficient to give a transverse spin-current, so we ignore the exchange interaction in this derivation. This considerably simplifies the work needed and allows us to find a completely analytical expression for the eigenfunctions. 

We start from the Landau-Lifshitz-Gilbert equation (LLG)
\begin{equation}
\partial_t \nv (x, \rv,t) = \nv(x, \rv,t) \times \left[ \bm{H}_{\mathit{eff}} - \frac{\alpha}{S}\partial_t\nv(x, \rv,t) \right],
\end{equation}
where $\rv = (y, z)$. The classical ground state is
\begin{equation}
\hat{\bm{n}} = \frac{\expval{\nv}}{S} = \sin\phiM\, \xv + \cos\phiM\, \hat{\bm{z}},
\end{equation}
with the angle $\phiM$ is determined by Eq.~(\ref{eq:angle-magnetic-field}) in the main text.
We write the solution to the LLG as fluctuations on this ground state with
\begin{equation}
\psi=\frac{1}{S\sqrt{2}}\left(\av+i\bv\right)\cdot\nv (\rv,t),
\end{equation}
where $\av,\bv$ are orthogonal unit vectors chosen such that by $\av \times \bv=\hat{\bm{n}}$.
The effective magnetic field is given by
\begin{equation}
\bm{H}_{\mathit{eff}} = \bm{H} + \bm{H}_D;\quad \bm{H}_D = \bm{H}_D^{(0)} + \nabla\chi,
\end{equation}
where $\bm{H}_D$ is the dipolar field with a static component $\bm{H}_D^{(0)}$ and a dynamic component, $\nabla\chi$. $\bm{H}$ is the external field.
We transform to Fourier space with the relations
\begin{equation}
\psi\left(x, \rv,\omega\right)  =\int\frac{d^{2}k}{\left(2\pi\right)^{2}}e^{i\kv\cdot\rv}\psi\left(x,\kv,\omega\right),\quad
\chi\left(x, \rv,\omega\right)  =\int\frac{d^{2}k}{\left(2\pi\right)^{2}}e^{i\kv\cdot\rv}\chi\left(x,\kv,\omega\right).
\end{equation}
We only consider the situation where $k_y=0$, so $\kv=k \hat{\bm{z}}$. Outside the film the dynamics of the dipolar field are governed by
\begin{equation}
\left(-k^{2}+\partial_{x}^{2}\right)\chi\left(x,\kv,\omega\right)=0,\quad x\geq\frac{d}{2}
\end{equation}
which has solutions
\begin{equation}
\chi\left(x,\kv,\omega\right)=\begin{cases}
\chi\left(\frac{d}{2},\kv,\omega\right)e^{-|k|\left(x-d/2\right)}, & x\geq\frac{d}{2};\\
\chi\left(-\frac{d}{2},\kv,\omega\right)e^{|k|\left(x+d/2\right)}, & x\leq\frac{d}{2}.
\end{cases}
\end{equation}
The boundary conditions for $\chi$ at the top and bottom of the thin film are 
\begin{multline}
\qquad\partial_{x}\chi\left(x,\kv,\omega\right)\Bigr\vert_{x=\pm\frac{d}{2}\mp0^{+}}
+4\pi M_{S}\frac{1}{\sqrt{2}} \Bigg[\xv\cdot\left(\av-i\bv\right)\psi\left(\pm\frac{d}{2},\kv,\omega\right)\\
+\xv\cdot\left(\av+i\bv\right)\psi^{*}\left(\pm\frac{d}{2},-\kv,-\omega\right)\Bigg]
=\mp |k| \chi\left(\pm\frac{d}{2},\kv,\omega\right)
\label{eq:bc}
\end{multline}
and the bulk equation of motion for $\chi$ is
\begin{multline}
\qquad \left(-k^{2}+\partial_{x}^{2}\right)\chi\left(x,\kv,\omega\right)
+4\pi M_{S}\frac{1}{\sqrt{2}}
\Big[
\left(\av-i\bv\right)\cdot\left(i\kv+\xv\partial_{x}\right)\psi\left(x,\kv,\omega\right)
\\+\left(\av+i\bv\right)\cdot\left(i\kv+\xv\partial_{x}\right)\psi^{*}\left(x,-\kv,-\omega\right)
\Big]
= 0, \quad\left|x\right|\leq\frac{d}{2}.
\end{multline}
For the magnon field we have the bulk equation of motion
\begin{equation}
\left[\left(1+i\alpha\right)\omega-\boldsymbol{H}\cdot\expn-4\pi M_{S}\left(\xv\cdot\expn\right)^{2}\right]\psi\left(x,\kv,\omega\right)+h_{D}\left(x,\kv,\omega\right)=0.
\end{equation}
This gives the solution 
\begin{equation}
\psi\left(x,\kv,\omega\right) = G\left(\omega\right)h_{D}\left(x,\kv,\omega\right),
\end{equation}
where 
\begin{equation}
G\left(\omega\right)=\left[-\left(1+i\alpha\right)\omega+\boldsymbol{H}\cdot\expn+4\pi M_{S}\left(\xv\cdot\expn\right)^{2}\right]^{-1}.
\end{equation}
For brevity we define
\begin{equation}
\Delta G\left(\omega\right) \equiv G\left(\omega\right)+G^{*}\left(-\omega\right).
\end{equation}
From the bulk equation of motion the solution for the potential is
\begin{equation}
\chi\left(x,\kv,\omega\right)=\chi_{+}e^{qx}+\chi_{-}e^{-qx},
\end{equation}
where 
\begin{equation}
q = |k_z|\sqrt{\frac{a\left(\kv,\omega\right)}{b\left(\kv,\omega\right)}},
\end{equation}
with
\begin{align}
a\left(\kv,\omega\right) & =1+2\pi M_{s}\Delta G\left(\omega\right)\sin^{2}\phiM,\\
b\left(\kv,\omega\right) & =1+2\pi M_{S}\Delta G\left(\omega\right)\cos^{2}\phiM.
\end{align}
From the boundary conditions in Eq.~(\ref{eq:bc}) we then have the matrix equation
\begin{equation}
\begin{pmatrix}
\left(F_{+}\left(\kv,\omega\right)+\left|\kv\right|\right)e^{q\frac{d}{2}} & \left(F_{-}\left(\kv,\omega\right)+\left|\kv\right|\right)e^{-q\frac{d}{2}}\\
\left(F_{+}\left(\kv,\omega\right)-\left|\kv\right|\right)e^{-q\frac{d}{2}} & \left(F_{-}\left(\kv,\omega\right)-\left|\kv\right|\right)e^{q\frac{d}{2}}
\end{pmatrix}
\begin{pmatrix}
\chi_{+}\\
\chi_{-}
\end{pmatrix}
=
0
\end{equation}
where 
\begin{equation}
F_{\pm}\left(\kv,\omega\right)=-i\pi M_{S}\Delta G\left(
\omega \right)k_{z}\sin\left(2\phiM\right) \pm q\left(2\pi M_{S} \Delta G\left(
\omega \right)\cos^{2}\phiM+1\right).
\end{equation}

The solutions for the potential are then
\begin{equation}
\chi_{+}=-\chi_{-}\frac{\left(F_{-}\left(\kv,\omega\right)+\left|\kv\right|\right)}{\left(F_{+}\left(\kv,\omega\right)+\left|\kv\right|\right)}e^{-qd}
\end{equation}

which gives for the magnon field
\begin{multline}
\psi\left(x, \kv,\omega \right) = -G\left(\omega\right) 
\Bigg[ 
q \chi_{-} \cos\phiM 
\left( 
\frac{\left(F_{-}\left(\kv,\omega\right)+\left|\kv\right|\right)}{\left(F_{+}\left(\kv,\omega\right)+\left|\kv\right|\right)}
e^{qx-qd}
+e^{-qx} 
\right)
\\
+
i k_z \chi_{-}  \sin\phiM 
\left( 
-\frac{\left(F_{-}\left(\kv,\omega\right)+\left|\kv\right|\right)}{\left(F_{+}\left(\kv,\omega\right)+\left|\kv\right|\right)}
e^{qx-qd}
+\chi_{-}e^{-qx} 
\right) 
\Bigg].
\end{multline}

Because $F_\pm(\kv,\omega)$ depends linearly on $k_z \sin\left(2\phiM\right)$, the eigenfunctions for magnons travelling in $\pm k_z$ directions differ whenever $\sin\left(2\phiM\right) \neq 0$. This behaviour is in agreement with our numerics, which show that the difference between the transverse spin-current induced by left- and right-moving spin waves vanishes if the magnetization is either completely in- or out of plane. Ultimately the source of the linear term is therefore the boundary conditions in Eq.~(\ref{eq:bc}). Because the dipole-dipole interaction is a long-range interaction the boundary conditions interact with all the spin-waves in the thin film, carrying the inversion breaking at the interface. This thus allows a transverse spin current to flow.

\section{Dispersion}
\begin{figure*}
	\centering
	\includegraphics{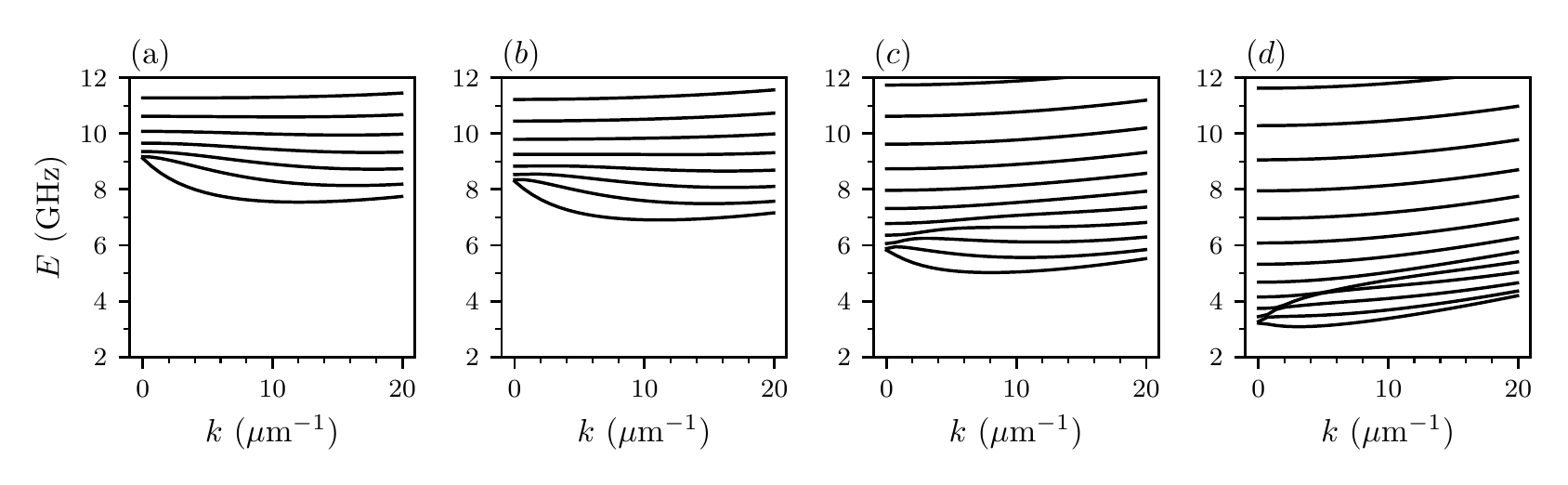}
	\caption{Spin wave dispersion of a YIG film with thickness $d=400a\approx \SI{0.48}{\mu m}$ for increasingly tilted magnetic field. The spin waves travel parallel to the in-plane projection of the magnetic field, such that $\kv=k \hat{\bm{z}}$. (a) $\phiH=\phiM=\ang{0}$, (b) $\phiH=\ang{30}, \phiM=\ang{18}$, (c) $\phiH=\ang{60}, \phiM=\ang{40}$ and (d) $\phiH=\ang{80}, \phiM=\ang{64}$.
		\label{fig:spin-wave-dispersion}}
\end{figure*}

We diagonalize the Hamiltonian in Eq.~(\ref{eq:d-matrix}) in the main text, in the absence of damping and spin pumping, from which we obtain the spin-wave energies \cite{colpaDiagonalizationQuadraticBoson1978}.
The spin-wave spectra are shown in Fig.~\ref{fig:spin-wave-dispersion} for multiple tilt angles of the magnetic field, for spin waves propagating parallel to the in-plane projection of the magnetic field, along the $k_z$ direction. The parameters used for these spectra are summarized in Table.~\ref{tab:values} in the main text. We show the regime of wavevectors where both dipole-dipole interactions and the exchange interaction are of roughly equal magnitude.
The exchange interaction dominates for large wavevectors and gives a quadratic wavevector dependence, curving the bands upwards. For small wavevector the dipole-dipole interaction is the dominant term in the Hamiltonian, which suppresses the quadratic behavior. 
Comparing these dispersion with both the numerical and experimental results \cite{bozhkoUnconventionalSpinCurrents2020} the general shape of the dispersions matches well, and the same shift down in energy is observed as the magnetic field is tilted.

\section{Complete amplitude factors}
The amplitude factors in Eq.~(\ref{eq:d-matrix}) in the main text are
\begin{align}
A_{\kv}(x_{ij}) & =\sum_{\rv_{ij}}e^{-i\kv\cdot\rv}A(x_{i}-x_{j},\rv), \nonumber \\
& =\delta_{ij}\left[\cos\left(\phiH-\phiM\right)h+S\sum_{n}\left(\sin^{2}\phiM D_{0}^{xx}(x_{in})+\cos^{2}\phiM D_{0}^{zz}(x_{in})+\sin\phiM\cos\phiM D_{0}^{xz}(x_{in})\right)\right] \nonumber \\
&\quad -\frac{S}{2}\left[\cos^{2}\phiM D_{\kv}^{xx}(x_{ij})+D_{\kv}^{yy}(x_{ij})+\sin^{2}\phiM D_{\kv}^{zz}(x_{ij})-2\sin\phiM\cos\phiM D_{\kv}^{xz}(x_{ij})\right] + SJ_{\kv}(x_{ij}),\\
B_{\kv}(x_{ij}) & =\sum_{\rv_{ij}}e^{-i\kv\cdot\rv}B(x_{i}-x_{j},\rv), \nonumber \\
& =-\frac{S}{2}\Big[\cos^{2}\phiM D_{\kv}^{xx}(x_{ij})-D_{\kv}^{yy}(x_{ij})+\sin^{2}\phiM D_{\kv}^{zz}(x_{ij})-\cos\phiM\sin\phiM D_{\kv}^{xz}(x_{ij}) \nonumber\\
&\qquad +i\sin\phiM D_{\kv}^{yz}(x_{ij})-i\cos\phiM D_{\kv}^{xy}(x_{ij})\Big],
\end{align}
where
\begin{equation}
J_{\kv}(x_{ij}) = J\left[\delta_{ij} \left( 6 - \delta_{j1} -\delta_{jN} -2\cos(k_ya)-2\cos(k_za) \right) - \delta_{ij+1} - \delta_{ij-1} \right]
\end{equation}
and $\rv_{ij} = (y_{ij}, z_{ij})$.

The dipole-dipole interaction is written as a tensor
\begin{equation}
D^{\alpha\beta}_{\kv} (x_{ij}) = \sum_{\rv_{ij}} e^{-i\kv\cdot\rv_{ij}} D_{ij}^{\alpha\beta},
\label{eq:dipole-sums}
\end{equation}
where
\begin{equation}
D_{ij}^{\alpha\beta} = \mu^2(1-\delta_{ij}) \frac{\partial^2}{\partial R_{ij}^\alpha \partial R_{ij}^\beta} \frac{1}{|\Rv_{ij}|}. \label{eq:dipole}
\end{equation}

For small wavevectors the sums in Eq.~(\ref{eq:dipole-sums}) are slowly converging, so we use the Ewald summation method as outlined by \textcite{kreiselMicroscopicSpinwaveTheory2009}. With this method the sums are split in two parts: one sum over real space and a one sum over reciprocal space. These sums are much faster to converge. We first write the sums as a derivative of 
\begin{equation}
I_{\kv}(x_{ij})=\mu^{2}\sum_{y_{ij},z_{ij}}\frac{e^{-i(k_{y}y_{ij}+k_{z}z_{ij})}}{(x_{ij}^{2}+y_{ij}^{2}+z_{ij}^{2})^{5/2}},
\end{equation}
such that we have 
\begin{align}
D_{\kv}^{xx} & =\left[\frac{\partial^{2}}{\partial k_{z}^{2}}+\frac{\partial^{2}}{\partial k_{y}^{2}}+2x_{ij}^{2}\right]I_{\kv}(x_{ij}),\\
D_{\kv}^{yy} & =\left[\frac{\partial^{2}}{\partial k_{z}^{2}}-2\frac{\partial^{2}}{\partial k_{y}^{2}}-x_{ij}^{2}\right]I_{\kv}(x_{ij}),\\
D_{\kv}^{zz} & =\left[\frac{\partial^{2}}{\partial k_{y}^{2}}-2\frac{\partial^{2}}{\partial k_{z}^{2}}-x_{ij}^{2}\right]I_{\kv}(x_{ij}),\\
D_{\kv}^{xy} & =3ix_{ij}\frac{\partial}{\partial k_{y}}I_{\kv}(x_{ij}),\\
D_{\kv}^{xz} & =3ix_{ij}\frac{\partial}{\partial k_{z}}I_{\kv}(x_{ij}),\\
D_{\kv}^{yz} & =3\frac{\partial}{\partial k_{z}\partial k_{y}}I_{\kv}(x_{ij}).
\end{align}
Note the symmetries $D_{\kv}^{yy}=D_{\kv}^{zz}(k_{y}\rightarrow k_{z},k_{z}\rightarrow k_{y})$ and $D_{\kv}^{xz}=D_{\kv}^{xy}(k_{y}\rightarrow k_{z},k_{z}\rightarrow k_{y})$, so we need not derive the full form of all dipolar sums.
Then, after applying the Ewald summation, we have
\begin{align}
D_{\kv}^{xx}(x_{ij})&=\frac{\pi\mu^{2}}{a^{2}}\sum_{\gv}
\left(\frac{8\sqrt{\varepsilon}}{3\sqrt{\pi}}e^{-p^{2}-q^{2}}-|\kv+\gv|f(p,q)\right) \nonumber\\
&\qquad -\frac{4\mu^{2}}{3}\sqrt{\frac{\varepsilon^{5}}{\pi}}\sum_{\rv}\left(|\rv_{ij}|^{2}-3x_{ij}^{2}\right)\cos\left(k_{y}y_{ij}\right)\cos\left(k_{z}z_{ij}\right)\varphi_{3/2}(|\rv_{ij}|^{2}\varepsilon),\\
D_{\kv}^{yy}(x_{ij})&=\frac{\pi\mu^{2}}{a^{2}}\sum_{\gv}
\left(\frac{4\sqrt{\varepsilon}}{3\sqrt{\pi}}e^{-p^{2}-q^{2}}-\frac{(k_{y}+g_{y})^{2}}{|\kv+\gv|}f(p,q)\right)  \nonumber	 \\
&\qquad -\frac{4\mu^{2}}{3}\sqrt{\frac{\varepsilon^{5}}{\pi}}\sum_{\rv}\left(|\rv_{ij}|^{2}-3y_{ij}^{2}\right)\cos\left(k_{y}y_{ij}\right)\cos\left(k_{z}z_{ij}\right)\varphi_{3/2}(|\rv_{ij}|^{2}\varepsilon), \\
D_{\kv}^{xy}(x_{ij}) &= i\frac{\pi\mu^2}{a^2} \sig(x_{ij})\sum_{\gv} (k_y + g_y)f(p,q)  \nonumber  \\
&\qquad + i\frac{4\varepsilon^{5/2}\mu^2	}{\sqrt{\pi}}x_{ij} \sum_{\rv} \sin(k_y y_{ij})\cos(k_z z_{ij}) \varphi_{3/2}(|\rv_{ij}|^{2}\varepsilon),\\
D_{\kv}^{yz} (x_{ij}) &= -\frac{\pi \mu^2} {a^2} \sum_{\gv}
\frac{(k_y + g_y) (k_z + g_z)}{|\kv+\gv|}  f(p, q) \nonumber\\
&\qquad+4 \frac{\varepsilon^{5 / 2} \mu^2}{\sqrt{\pi}} \sum_{\rv}  y_{ij} z_{ij} 
\sin(k_y  y_{ij})  sin(k_z  z_{ij})  \varphi_{3/2}(|\rv_{ij}|^{2}\varepsilon),
\end{align}
where 
\begin{equation}
\varphi_{3/2}(x) = e^{-x} \frac{3 + 2x}{2x^2} + \frac{3\sqrt{\pi} \erfc \left( \sqrt{x} \right)}{4x^{5/2}}
\end{equation}
and $q=x_{ij}\sqrt{\varepsilon}$, $p=|\kv + \bm{g}|/(2\sqrt{\varepsilon})$ and
$f(p,q)=e^{-2pq}\erfc(p-q) + e^{2pq}\erfc(p+q)$.
The sums are either over the real space lattice or the reciprocal lattice, where the reciprocal lattice vectors are $g_y=2\pi m$, $g_z=2\pi n$, $\{m,n\}\in \mathbb{Z}$.
$\varepsilon$ determines the ratio between the reciprocal and real sums. We choose $\varepsilon=a^{-2}$, such that $2pq \approx 1$ and $\exp[\pm 2pq]$ converges quickly.

\section{Current contributions}
In the continuity equation for the angular momentum in the main text, Eq.~(\ref{eq:continuity}), the explicit form of the terms is
\begin{align}
I_{i}^{\alpha}(\kv, \omega) &= 2\alpha_i\Im\left[b_{\kv}^* (x_i) \partial_t b_{\kv} (x_i)\right] \\
I_{i}^{h}(\kv, \omega) &= -\sqrt{2S}\Im\left[h_{i}b_{\boldsymbol{k}}^{*}\left(x_{i}\right)\right],\\
I^{ex}_{i\rightarrow j}(\kv, \omega)&=i\left(1-\delta_{ij}\right)S J_{\kv}\left( x_{ij} \right) b_{\boldsymbol{k}}^{*}\left(x_{i}\right)b_{\boldsymbol{k}}\left(x_{j}\right).\\
I^{dip-dip}_{i\rightarrow j}(\kv, \omega)&=i\Bigg[\left(1-\delta_{ij}\right) A_{\boldsymbol{k}}^{dip}\left(x_{ij}\right) b_{\boldsymbol{k}}^{*}\left(x_{i}\right)b_{\boldsymbol{k}}\left(x_{j}\right) \nonumber \\
&\qquad -\frac{B_{\boldsymbol{k}}\left(x_{ij}\right)}{2}b_{-\boldsymbol{k}}\left(x_{i}\right)b_{\boldsymbol{k}}\left(x_{j}\right)  +\frac{B_{\boldsymbol{k}}^{*}\left(x_{ij}\right)}{2}b_{\boldsymbol{k}}^{*}\left(x_{i}\right)b_{-\boldsymbol{k}}^{*}\left(x_{j}\right)\Bigg],
\end{align}
where $A_{\boldsymbol{k}}^{dip}\left(x_{ij}\right)=A_{\boldsymbol{k}}^{h=J=0}\left(x_{ij}\right)$, i.e., only the contributions from the dipole-dipole interaction. Note that $B_{\kv}\left( x_{ij}\right)$ already includes only dipole-dipole interactions.

\end{document}